\documentclass[preprint,12pt]{elsarticle}

\usepackage[utf8]{inputenc}
\usepackage{graphicx}
\usepackage{amsmath}
\usepackage{amssymb}
\usepackage{booktabs}
\usepackage{multirow}
\usepackage{url}
\usepackage{float}
\usepackage{hyperref}
\usepackage{xcolor}
\hypersetup{
    pdftitle={A Resilience Evaluation Framework for Electric Distribution Systems: Historical Weather Conditioning, Sensitivity Analysis, and a Flooding-Aware Extension},
    pdfauthor={Xuesong Wang, Caisheng Wang, Carol Miller, Amir Shahin Kamjou, John Norton}
}

\journal{International Journal of Electrical Power \& Energy Systems}

\newif\ifanonymous
\anonymousfalse

\begin{document}

\begin{frontmatter}

\title{A Resilience Evaluation Framework for Electric Distribution Systems: Historical Weather Conditioning, Sensitivity Analysis, and a Flooding-Aware Extension}

\ifanonymous
\else
\author[1]{Xuesong Wang}
\ead{xswang@wayne.edu}
\author[1]{Caisheng Wang\corref{cor1}}
\ead{cwang@wayne.edu}
\author[2]{Carol Miller}
\ead{ab1421@wayne.edu}
\author[2]{Amir Shahin Kamjou}
\ead{Amir.Kamjou@wayne.edu}
\author[3]{John Norton}
\ead{john.norton@glwater.org}

\cortext[cor1]{Corresponding author}

\affiliation[1]{organization={Department of Electrical and Computer Engineering, Wayne State University},
            addressline={42 W. Warren Ave.},
            city={Detroit},
            postcode={48201},
            state={MI},
            country={USA}}
\affiliation[2]{organization={Department of Civil and Environmental Engineering, Wayne State University},
            addressline={42 W. Warren Ave.},
            city={Detroit},
            postcode={48201},
            state={MI},
            country={USA}}
\affiliation[3]{organization={Great Lakes Water Authority},
            addressline={735 Randolph, Suite 1900},
            city={Detroit},
            postcode={48226},
            state={MI},
            country={USA}}
\fi

\begin{abstract}
Evaluating resilience in electric distribution systems under severe weather requires models that can connect network topology, hazard simulation, fragility modeling, restoration assumptions, repair strategy, and downstream consequences. This paper extends our prior graph-based resilience evaluation framework for power distribution systems in three ways: it adds analysis conditioned on historical events with real outage and weather data, introduces sensitivity studies for key modeling assumptions, and includes a coupled power-flooding extension for sewage-backup assessment. Historical wind events drive Monte Carlo simulations conditioned on real weather, and the observed outage trajectories are treated as realized historical samples for comparison. Wind-event resilience metrics stabilize at approximately 256 episodes, and outage peak, duration, and outage intensity change systematically with fragility parameters, network topology, restoration assumptions, and repair strategies. In a separate 1000-episode joint power-flooding simulation, episodes with at least one flooded customer occur in 1.9\% of episodes overall, and both flood occurrence and flood intensity increase with outage intensity, showing a selective power-to-flood consequence pathway. Overall, the framework provides a practical basis for resilience assessment, comparative scenario analysis, and coupled power-flooding studies in a limited public-data setting, while also suggesting that more detailed utility data could further improve simulation realism.

\end{abstract}

\begin{keyword}
Electric distribution system resilience \sep Flooding-aware resilience assessment \sep Fragility modeling \sep Historical weather conditioning \sep Monte Carlo simulation \sep Sensitivity analysis
\end{keyword}

\end{frontmatter}

\section{Introduction}
\label{sec:introduction}
Extreme weather events continue to expose critical vulnerabilities in electric distribution systems, where prolonged outages can disrupt households, public services, and local economic activity \cite{Younesi_2022,Potts_2024}. These impacts matter not only to utilities, but also to public agencies and communities responsible for infrastructure investment, emergency planning, and service continuity \cite{Potts_2024,Brelsford_2024}. The problem is broader than electricity alone: outages can propagate into other interdependent community systems, including water and wastewater services \cite{Oikonomou_2021}. As a result, resilience assessment needs to support both engineering analysis and broader community planning.

Resilience assessment is also difficult because it focuses on rare, high-impact events \cite{Mishra_2021,Younesi_2022}. Severe storms, ice events, and compound disruptions do not occur often enough to yield abundant local observations, and the outage records that are publicly available are usually incomplete, spatially coarse, or aggregated for other purposes \cite{Brelsford_2024,Lee_2024}. This makes resilience fundamentally different from routine reliability analysis: the events of greatest interest are precisely those for which detailed outage and restoration data are hardest to obtain from public sources \cite{Brelsford_2024,Lee_2024}. For this reason, resilience studies cannot rely on historical observations alone. They also need simulation frameworks that can explore plausible outage and recovery behavior under different assumptions \cite{Cresta_2021}.

From an engineering perspective, such simulation requires more than a weather-exposure indicator. It requires explicit representation of network topology, weather simulation, component fragility, restoration timing, and crew-based repair decisions \cite{Cresta_2021,Zhang_2023}. Existing work has advanced resilience assessment through statistical outage analysis, weather-outage mapping, predictive models, and simulation-based planning \cite{Mishra_2021,Lee_2024,Prieto_Godino_2025,Cresta_2021,Zhang_2023}. Much of this literature, however, still emphasizes aggregated outage datasets, outage prediction, restoration support, or single-infrastructure simulation, rather than historically conditioned engineering simulation with sensitivity studies and coupled local consequences \cite{Lee_2024,Prieto_Godino_2025,Cresta_2021,Oikonomou_2021}.

Our prior conference paper introduced a graph-based simulation framework for power resilience estimation and enhancement \cite{Wang_2025}. That work used a synthetic distribution network informed by real data, synthetic wind hazards, and Monte Carlo simulation to estimate resilience and study distributed energy resource (DER)-based enhancement on a large graph. The conference paper established the core framework architecture, but it did not yet connect the framework to historical outage observations, evaluate studies conditioned on specific events, examine sensitivity in a calibration-oriented way, or couple the power model to another community infrastructure system.

The present study addresses those next steps by using collected utility outage data and real weather data to compare simulations conditioned on historical events with observed outage behavior in a calibration-oriented rather than deterministic-replay sense. It also examines sensitivity to fragility, topology, restoration, and repair assumptions, and extends the framework to a joint power-flooding setting in which pump outages can trigger sewage-backup flooding. The remainder of the paper is organized as follows. Section~\ref{sec:related} reviews related work. Section~\ref{sec:method} presents the framework, the weather analysis conditioned on historical events, and the coupled power-flooding extension. Section~\ref{sec:experiments} presents the experiments and results. Section~\ref{sec:discussion} discusses the main findings and limitations. Section~\ref{sec:conclusion} concludes the paper.

\section{Related Work}
\label{sec:related}
\subsection{Electric distribution system resilience assessment}
Recent review papers describe electric distribution system resilience as a combined planning, operation, and recovery problem that spans network hardening, distributed energy resources, reconfiguration, and restoration under high-impact, low-probability events \cite{Mahdi_2025,Younesi_2022,Tang_2024}. At the framework level, recent studies have proposed simulation-oriented models for electric distribution grid resilience assessment \cite{Cresta_2021}. Together, these studies establish the need for distribution-level resilience modeling, but they also show that the literature remains split across review papers, planning formulations, and standalone simulation studies.

Our work is closest to the simulation-oriented line, but differs in emphasis. Rather than proposing a new resilience index or a new planning objective, we focus on a graph-based Monte Carlo framework that can be conditioned on historical events, exercised under alternative modeling assumptions, and extended to downstream consequence analysis.

\subsection{Weather data, outage data, and empirical resilience context}
Publicly available outage and weather data have improved the empirical study of extreme-event impacts on power systems. County-scale outage records now provide a multi-year dataset for U.S. outage-event analysis \cite{Brelsford_2024}, and recent work has combined weather and outage data to quantify resilience across the U.S. power grid \cite{Lee_2024}. Event-focused studies have also used observed outage records to interpret resilience performance under severe storms such as Winter Storm Uri \cite{Potts_2024}. In parallel, data-driven outage prediction models continue to advance, including recent deep-learning approaches for weather-related outage prediction in large-scale distribution grids \cite{Prieto_Godino_2025,Wang_2024}.

These studies strengthen empirical benchmarking, descriptive resilience analysis, and predictive outage modeling. However, they do not by themselves provide a graph-level outage-and-restoration simulator that can test how alternative fragility parameters, network topology, and repair assumptions change event consequences under the same historical forcing. That gap motivates the simulation perspective conditioned on historical events that is used in this paper.

\subsection{Fragility, environmental exposure, and restoration modeling}
Recent outage studies have shown that weather exposure and spatial heterogeneity matter when modeling outage behavior in large distribution systems \cite{Prieto_Godino_2025,Lee_2024}. These findings support the need to expose fragility assumptions explicitly rather than treating outage rates as fixed black-box parameters.

On the restoration side, recent work continues to emphasize the importance of crew dispatch, switching, and mobile resource coordination in post-event service recovery \cite{Zhang_2023,Tang_2024}. Much of this literature is optimization-oriented. The restoration model used here remains comparatively lightweight so that fragility, topology, and repair-strategy assumptions can be compared systematically over large Monte Carlo ensembles under common event inputs.

\subsection{Interdependent infrastructure and flooding consequences}
Recent interdependence studies have highlighted the need to move beyond single-network analysis when evaluating infrastructure resilience. Reviews of interdependent water and power systems show that recovery processes and cross-system dependencies can substantially alter consequence propagation \cite{Oikonomou_2021}. Flood-focused infrastructure studies also demonstrate that geographically correlated flood disruptions can create network stress patterns that are not visible in single-asset analyses \cite{Raman_2022}. These observations are directly relevant to urban settings such as Detroit, where power outages and sewage-backup flooding can interact at the neighborhood scale.

Relative to this literature, this study emphasizes a simulation workflow conditioned on historical events that remains tractable in a limited public-data setting. We extend our prior graph-based framework for power resilience to support weather-event assessment, structured sensitivity analysis, and a coupled power-flooding consequence layer, so that the same simulation workflow can be used to study both outage impacts and a coupled community-level consequence pathway.

\section{Method}
\label{sec:method}
\subsection{Overview}
The workflow used in this paper has four stages: (i) curate historical outage events from utility outage records, (ii) characterize each event using High-Resolution Rapid Refresh (HRRR) \cite{Dowell_2022,James_2022} weather fields, (iii) run a stochastic outage-and-restoration simulation on the graph-based distribution model conditioned on that historical weather event, and (iv) assess the resulting outage distribution against observed outage summaries and coarse external comparisons.

\subsection{Graph-based electric power system model}
The underlying simulation framework was introduced in our conference paper \cite{Wang_2025}. The electric distribution system is represented as a graph whose nodes and lines carry the attributes needed for resilience analysis, including feeder identity, customer counts, weather-patch assignment, vegetation exposure, and underground versus overhead status. In the historically conditioned setting studied here, the hazard field is no longer drawn from a synthetic storm generator. Instead, it is taken from historical weather inputs for a specific event, derived from HRRR, and mapped onto the study-area patches used by the graph model.

Each Monte Carlo episode applies the same event-specific weather forcing to the graph, samples weather-driven component failures, propagates service interruption through the network, and then applies crew-based restoration logic to produce an outage trajectory. When the sewage network is enabled, pump outages induced by electrical service loss also trigger the coupled power-flooding extension described below.

\begin{figure}[htbp]
\centering
\includegraphics[width=0.9\textwidth]{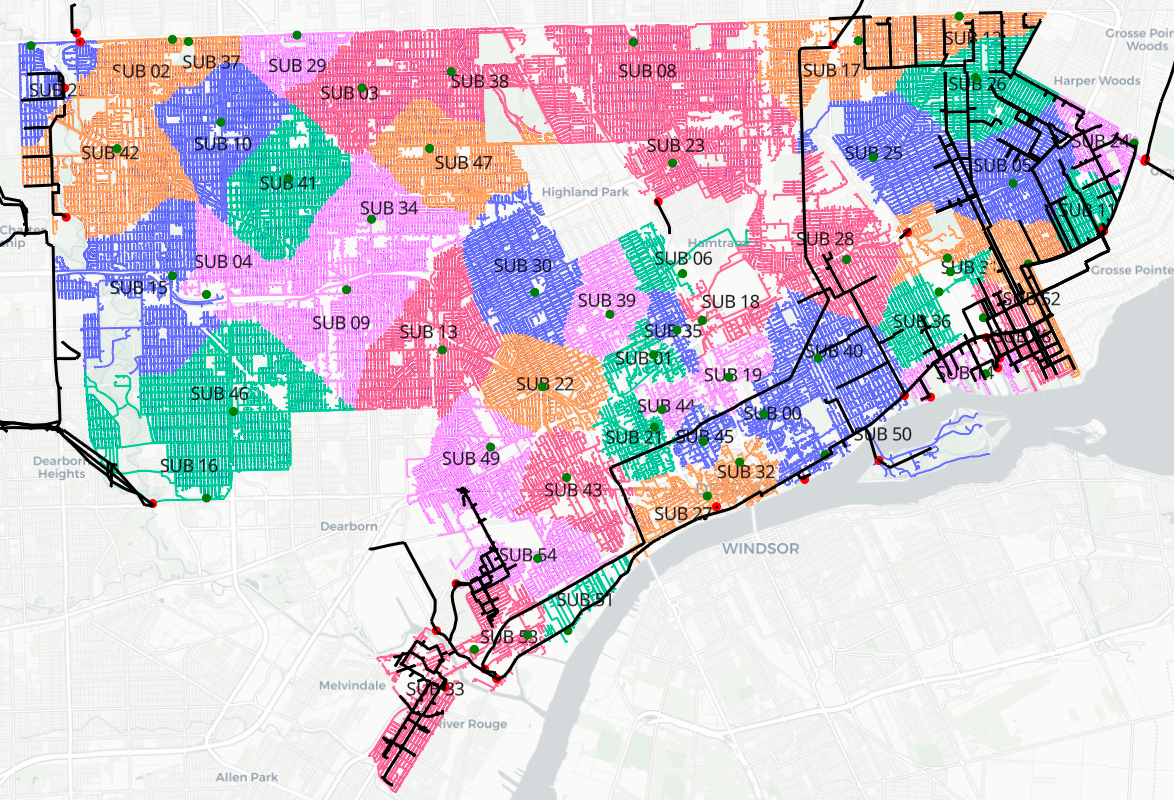}
\caption{Detroit study-area network topology used by the simulation framework, with feeder service areas and the overlaid sewage network shown in black.}
\label{fig:detroit_topology_service_areas}
\end{figure}

\subsection{Historical outage curation and event definition}
Historical outage polygons are intersected with the customer locations represented in the power graph, and the resulting graph-derived customer counts are treated as the outage proxy. This avoids reliance on utility polygon attributes such as \texttt{NUM\_CUST}, which may be inconsistent across snapshots. Candidate event hours are identified when at least two feeders simultaneously exceed 5\% feeder outages. Nearby outage periods are merged across short gaps of up to three hours, and only merged events with duration of at least six hours are retained. Hours with near-systemwide outage coverage can also be excluded when they appear to reflect suspicious data artifacts rather than a physically plausible event.

The historical assessment set contains five wind-typed events after outage curation, weather review, and event filtering. Winter events were explored separately during model development, but they are not part of the main quantitative claims because the available study-period data contain too few curated winter outage events for defensible calibration, and the current public-data winter hazard assumptions remain insufficiently constrained.

\subsection{Weather forcing and event typing}
For each curated event, hourly HRRR data are assembled over the event weather window. HRRR is used because it provides hourly, convection-allowing weather analyses and forecast products suitable for regional hazard reconstruction \cite{Dowell_2022,James_2022}. Event typing begins with threshold-based characterization over the study footprint. In this study, the main quantitative analysis focuses on wind-relevant events, identified when the spatial p95 wind gust reaches at least 17~m/s or the spatial maximum gust reaches at least 22~m/s for at least two hours.

\begin{figure}[htbp]
\centering
\includegraphics[width=0.7\textwidth]{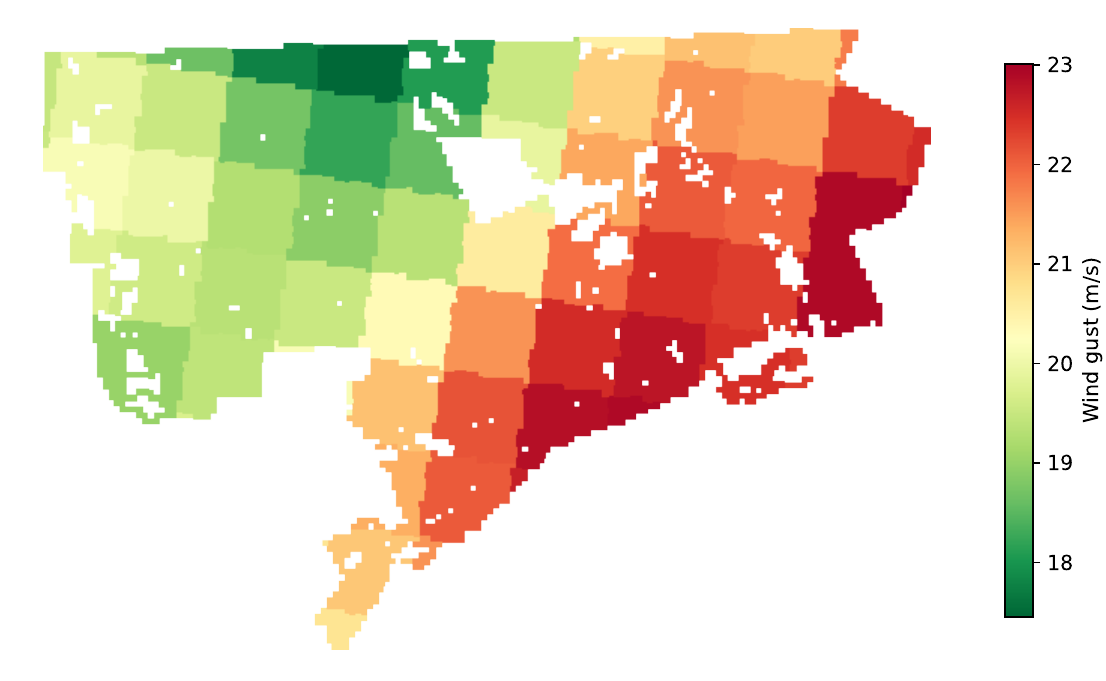}
\caption{Patch-level historical HRRR gust field for a representative curated wind event at its peak-gust hour, obtained by mapping HRRR weather fields onto the study-area patches over the Detroit footprint.}
\label{fig:historical_wind_field}
\end{figure}

The historical weather fields are mapped from HRRR grid cells to the study-area weather patches used by the simulation. Figure~\ref{fig:historical_wind_field} shows the peak-gust patch field for a representative curated wind event.

\subsection{Stochastic outage-and-restoration simulation}
Given an event and hazard class, the simulator executes repeated Monte Carlo episodes on the graph representation of the electric distribution system. All episodes for a given event share the same event-specific weather forcing. Stochasticity enters through probabilistic component failures and restoration decisions. The principal calibration parameter in the wind study is a fragility factor that scales the hazard-location thresholds used by the damage model. We also evaluate two other groups of assumptions that materially affect outage outcomes: network topology assumptions and operational assumptions. Here, service connections refer to the final customer-drop connections from the distribution network to premises. The topology ablation asks whether those connections are treated as underground or all overhead. The operational ablations consider restoration assumptions and repair strategy.

\subsection{Interpretation in a limited public-data setting}
The historical assessments developed here are not intended as deterministic event replay from public data alone. Several data limitations make exact reproduction unrealistic. First, outage severity is reconstructed by intersecting utility polygons with graph-defined customers, while utility feedback indicates that the polygons do not represent exact physical outage extent. Second, the current model begins systematic repair after the modeled hazard window, whereas real utilities may dispatch crews during ongoing weather when conditions permit. Third, the public proxy network omits many utility-internal details that materially affect outage distributions, including true feeder topology, underground versus overhead labeling, asset age and condition, protection settings, and operational restoration constraints. No utility-internal topology, asset-health, or restoration-log data are used in the current study. For that reason, each observed event is compared against a simulated distribution under the best available public assumptions, not treated as a point target for exact reproduction.

\subsection{Coupled power-flooding extension}
The coupled power-flooding extension links power outages to sewage-backup risk through the sewage pump network. When a sewage pump loses electrical service, it is treated as unavailable by the flooding model. Under normal operation, sewage flows downstream through the collection system and is lifted where pumps are installed. When a pump fails, however, sewage is no longer lifted along that downstream path, and surcharge is propagated upstream through the reversed sewage graph to represent backup flooding into contributing areas.

Once a pump has been out long enough to affect the upstream network, the corresponding conduit set is marked as flooded and assigned a flood radius that grows stochastically over time. In the current model, conduit buffers expand by a random hourly increment of approximately 30--60~m, with at least 100~m of upstream backup progression considered each hour. The union of the buffered flooded conduits defines the instantaneous flood polygon, and customer locations that fall within this evolving polygon are counted as flooded. When pumps recover, flood radii contract on conduit sets that are no longer sustained by still-failed pumps, allowing the flooded area to recede over time. Flood outcomes are summarized using flooded-customer peak, flood persistence duration, flooded-customer AUC (area under the curve), flooded-area peak, and flooded-area AUC.

\subsection{Assessment metrics}
For each event and model configuration, the simulator produces an empirical distribution of outage summaries from repeated Monte Carlo episodes under the same historical weather forcing. Let $\hat{F}_M(\cdot \mid W,\theta)$ denote the empirical distribution of summary metric $M$ for historical weather event $W$ under model configuration $\theta$, and let $M^{\mathrm{obs}}(W)$ denote the observed value derived from the outage polygons mapped onto the proxy graph. The calibration-oriented assessment therefore compares $M^{\mathrm{obs}}(W)$ against the simulated distribution $\hat{F}_M(\cdot \mid W,\theta)$ rather than asking the simulator to reproduce a single deterministic trajectory exactly.

For each event, we compare observed and simulated outage trajectories using three primary summary metrics:
\begin{itemize}
    \item outage peak customers,
    \item restoration duration (hours),
    \item outage intensity, defined here as the area under the outage curve (AUC).
\end{itemize}

Let $M^{\mathrm{obs}}$ and $M^{\mathrm{sim}}$ denote observed and simulated central estimates for a metric. We report ratio metrics
\begin{equation}
R_M = \frac{M^{\mathrm{sim}}}{M^{\mathrm{obs}}},
\end{equation}
with $R_M \approx 1$ indicating better alignment. In addition, we report two summary hit measures. A \emph{strict} hit requires the observed event metric to fall within the simulated p05--p95 interval. A \emph{pragmatic} hit is intentionally looser and asks whether the simulated mean is in the same broad level as the observation: within 0.5x to 2x of the observed value for each of the three primary outage metrics.

\subsection{Coarse external comparisons}
To triangulate realism beyond the distribution comparison conditioned on historical events, we run two coarse external comparisons at feeder-level aggregation:
\begin{itemize}
    \item outage-experience survey \cite{Tran_2026} versus simulated outage vulnerability.
    \item Detroit household flooding survey \cite{Larson_2021} versus simulated flood vulnerability.
\end{itemize}

These checks are treated only as coarse consistency signals. The survey datasets are legacy, based on broad self-reported past experience, and spatially privacy-shifted rather than event-aligned measurements. They are therefore not used as acceptance criteria for framework performance.

\section{Experiments and Results}
\label{sec:experiments}
\subsection{Experiment setup}
The main study conditioned on historical events uses five curated wind events and episode ladders from 32 to 1000 episodes so that convergence can be examined for the same three summary metrics used in the distribution comparison: outage peak, restoration duration, and outage intensity. Additional experiment groups evaluate wind-fragility sensitivity, network topology assumptions, restoration assumptions, and repair strategies. The coupled power-flooding extension is illustrated with a separate 1000-episode coupled wind run under the selected public-data wind configuration. Coarse external comparisons are based on feeder-level resilience outputs together with an outage-experience survey \cite{Tran_2026} and the Detroit household flooding survey \cite{Larson_2021}.

\subsection{Wind-event distribution behavior}
We first examine whether the simulation distributions conditioned on historical events become stable enough to support comparison with observed events. In this study, the relevant convergence target is not the observed utility event itself, but the simulated distribution produced under fixed historical weather forcing. The wind-event summary distributions stabilized substantially by approximately 256 episodes, which is why 1000 episodes were retained for the main wind study.

\begin{figure}[H]
\centering
\includegraphics[width=\textwidth]{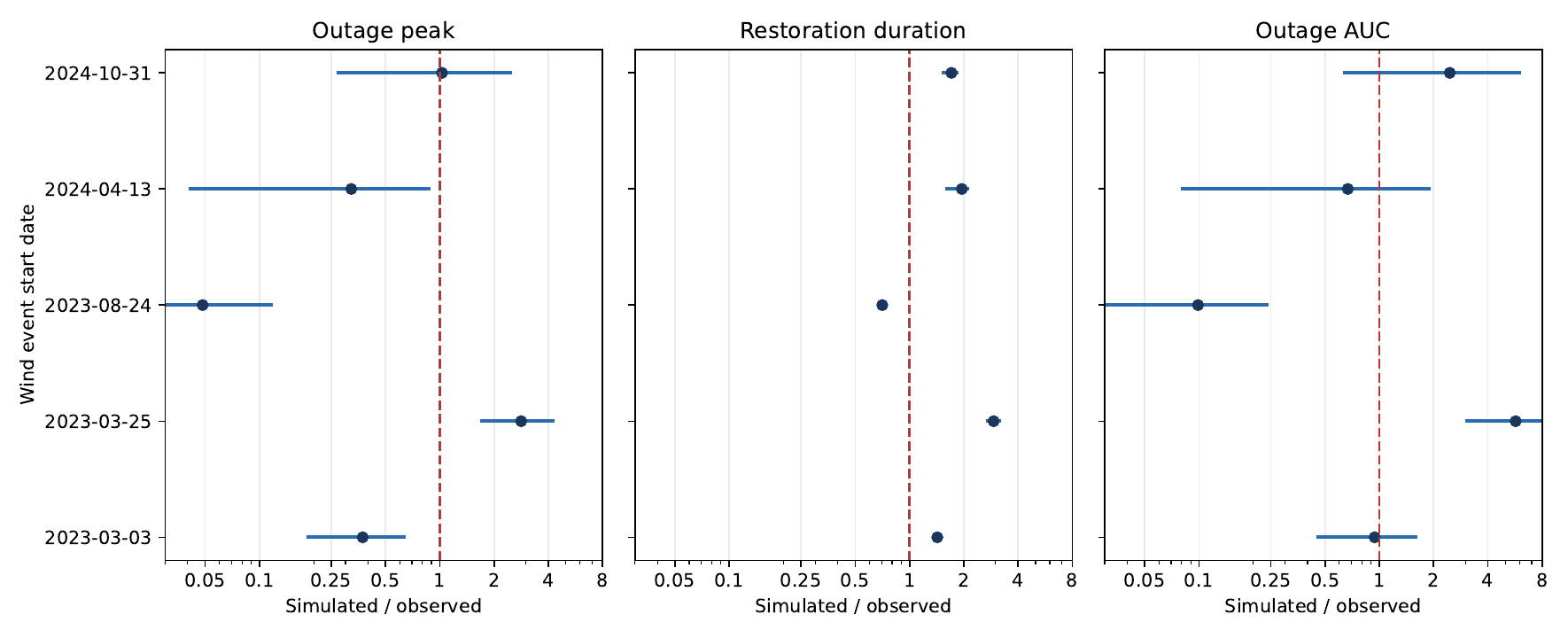}
\caption{Wind-event simulated distributions normalized by the observed event, shown as p05--p95 intervals and means for the three primary outage metrics.}
\label{fig:wind_interval_ratios}
\end{figure}

Figure~\ref{fig:wind_interval_ratios} shows the wind-event comparison directly. For each wind event and each of the three primary summary metrics, the figure reports the simulated p05--p95 interval and simulated mean after normalizing by the observed event value. The red dashed line at 1.0 corresponds to exact agreement with the observed sample. Across the 15 wind event-metric comparisons (five events times three metrics), the observed value falls inside the simulated p05--p95 band in 4 cases. Coverage is strongest for outage intensity (3/5 events), weaker for outage peak (1/5), and absent for restoration duration (0/5).

These counts are not interpreted as a validation score. Instead, they indicate which observed event attributes the present public-data model reproduces more readily than others. Under the strict p05--p95 coverage criterion, outage intensity shows the strongest agreement (3/5 events), while restoration duration shows no strict hits. Under the looser pragmatic criterion based on mean closeness within 0.5x to 2.0x of the observed value, restoration duration shows the strongest agreement (4/5 events), followed by outage intensity (2/5) and outage peak (1/5).

The retained wind events remain heterogeneous. For example, the March 3, 2023 event is close on outage intensity (ratio 0.94) but low on peak (0.37), the March 25, 2023 event is over-severe across all three metrics (peak ratio 2.83, duration ratio 2.93, intensity ratio 5.70), the August 24, 2023 summer event is strongly under-severe across all metrics (peak 0.05, duration 0.71, intensity 0.10), and the October 31, 2024 event is close on peak (1.03) but still long on duration (1.71) and outage intensity (2.46). The wind study is therefore interpreted as a comparative calibration exercise in a limited public-data setting.

\subsection{Sensitivity, network representation, and restoration assumptions}
\begin{figure}[ht]
\centering
\includegraphics[width=0.82\textwidth]{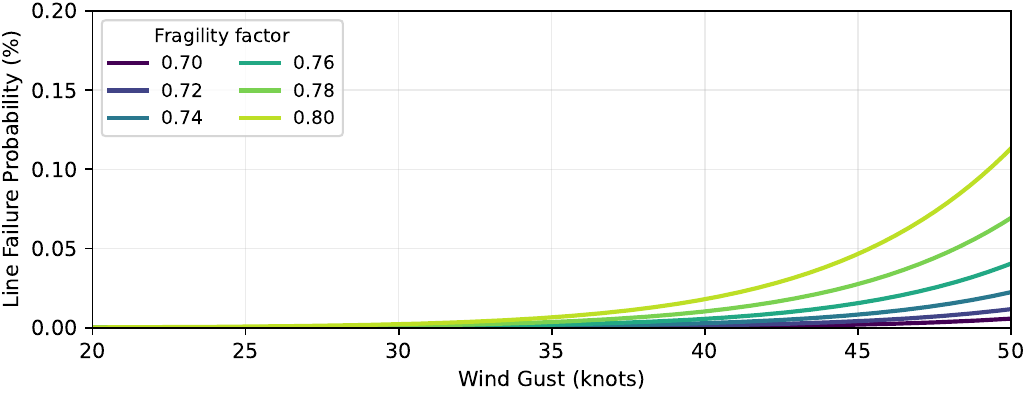}
\caption{Wind-fragility family used in the sensitivity study, with higher fragility factors shifting failure probability toward lower gust speeds.}
\label{fig:wind_fragility_family}
\end{figure}

\begin{figure}[ht]
\centering
\includegraphics[width=0.82\textwidth]{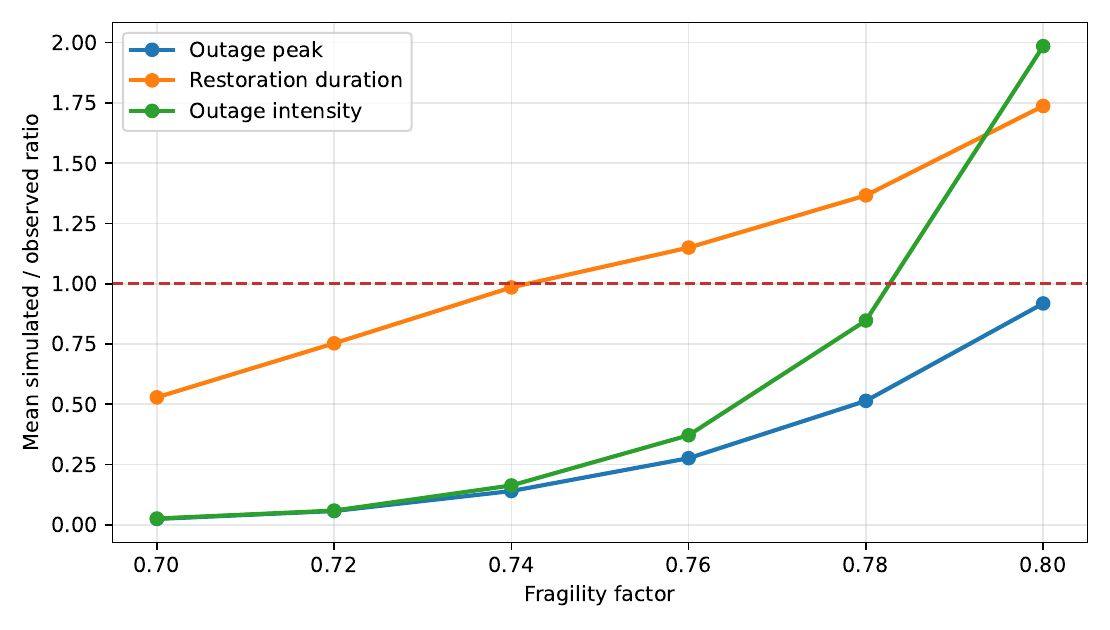}
\caption{Five-event wind sensitivity sweep showing monotone increases in mean simulated-to-observed ratios as the fragility factor increases.}
\label{fig:wind_fragility_response}
\end{figure}

The sensitivity studies are included to show that the framework responds coherently to modeling assumptions rather than producing a fixed outage regime. Figure~\ref{fig:wind_fragility_family} shows the family of wind-fragility curves used in the sensitivity study, and Figure~\ref{fig:wind_fragility_response} shows the corresponding five-event response in the primary outage metrics. Increasing the fragility factor shifts the failure-probability curve left by lowering the effective location threshold, and the sweep results show a monotone increase in mean simulated-to-observed ratios for outage peak, restoration duration, and outage intensity. Taken together, these figures show that the framework can be calibrated rather than treated as a black-box simulator with a single embedded damage level.

Table~\ref{tab:wind_ablation_compare} summarizes the topology, recovery-capacity, and repair-ordering comparisons using mean simulated-to-observed ratios across the five retained wind events.

\begin{table}[hb]
\centering
\caption{Topology, recovery-capacity, and repair-ordering comparisons for the five-event wind study.}
\label{tab:wind_ablation_compare}
\begin{tabular}{rrrr}
\toprule
Setup & PeakRatio & DurationRatio & AUCRatio \\
\midrule
1 & 0.920 & 1.743 & 1.975 \\
2 & 0.943 & 2.660 & 3.172 \\
3 & 0.920 & 2.788 & 3.375 \\
4 & 0.920 & 1.744 & 1.993 \\
5 & 0.920 & 1.743 & 1.149 \\
6 & 0.920 & 1.743 & 1.256 \\
\bottomrule
\end{tabular}

\end{table}

Setup 1 uses the service-underground graph, wind fragility factor 0.80, 12 repair teams, uniformly distributed repair times between 2 and 3 hours, and proximity-based repair ordering. Setup 2 changes only the topology, treating all service connections as overhead. Setup 3 changes only recovery capacity and repair times, using 6 repair teams and uniformly distributed repair times between 2 and 4 hours. Setups 4--6 keep the same topology, fragility, crew count, and repair-time setting as Setup 1, but replace proximity-based repair ordering with random, criticality-based, and hybrid-dynamic repair ordering, respectively. The hybrid-dynamic policy re-scores candidate repairs during restoration using a weighted combination of line criticality, crew travel distance, and remaining feeder backlog, with backlog priority increasing over recovery time.

These comparisons show that topology and recovery-capacity assumptions affect restoration duration and outage intensity much more strongly than outage peak. Relative to the reference setup, both the all-overhead topology assumption and the lower-capacity recovery setting substantially worsen duration and intensity outcomes. Across the repair-ordering comparison, random repair ordering produces the most severe outage-intensity ratio, the criticality-based policy the least severe, and the hybrid-dynamic policy an intermediate response. Under realistic deployment conditions, utility-specific topology and operational information are therefore likely to be as important as hazard intensity itself.

\subsection{Coupled power-flooding extension under wind episodes}
The coupled power-flooding extension is included to demonstrate that the same framework can propagate electrical-service disruptions into a second consequence layer with interpretable behavior. In the 1000-episode simulation, episodes with any flooded customers occur in 19 of 1000 episodes (1.9\%), and episodes with any flooded area occur in 27 of 1000 episodes (2.7\%). Flooding is therefore not a ubiquitous byproduct of every outage episode.

\begin{figure}[H]
\centering
\includegraphics[width=0.88\textwidth]{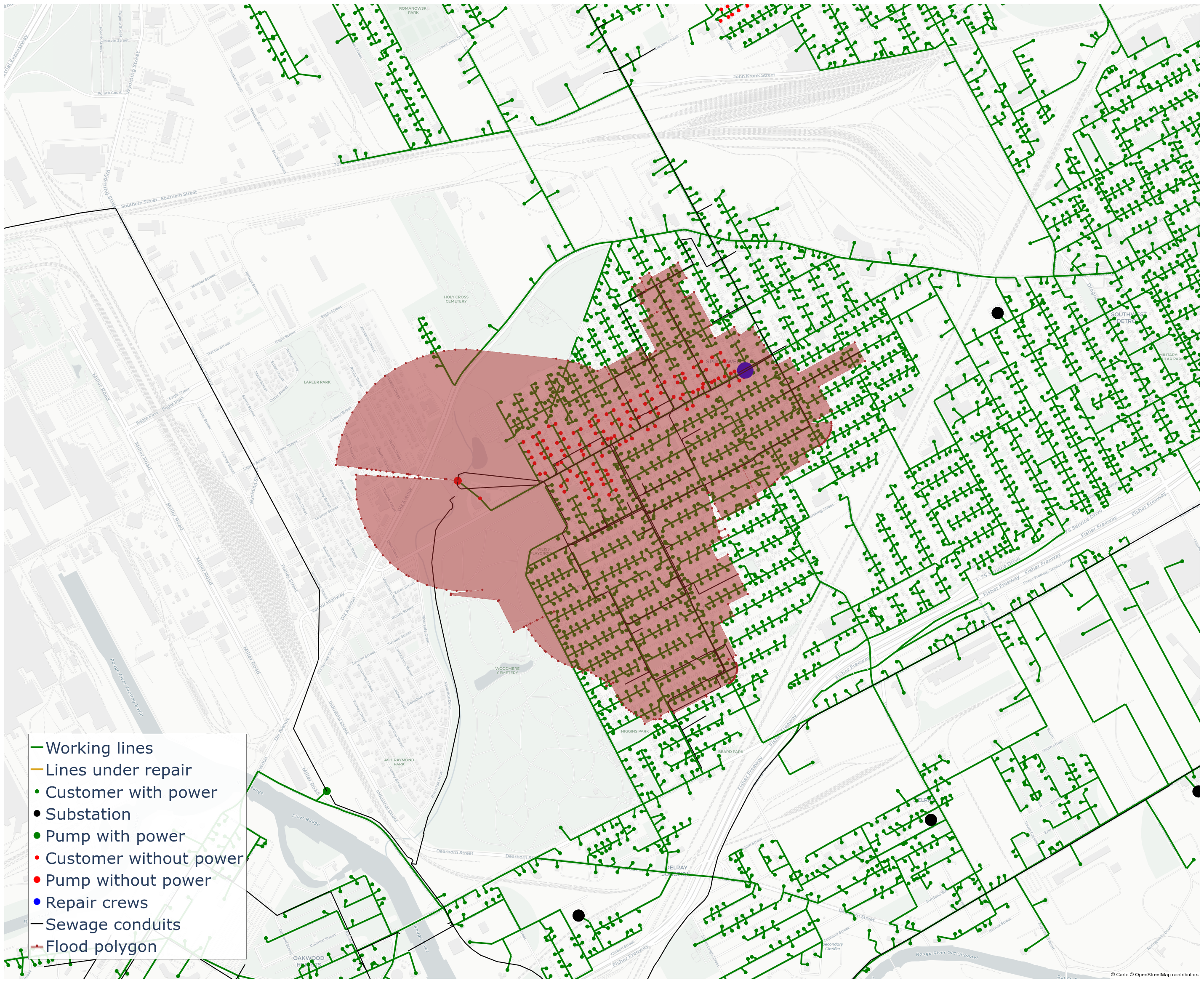}
\caption{Representative flooded-area footprint from the coupled wind run generated by the coupled power-flooding extension.}
\label{fig:flood_area_snapshot}
\end{figure}

\begin{figure}[H]
\centering
\includegraphics[width=\textwidth]{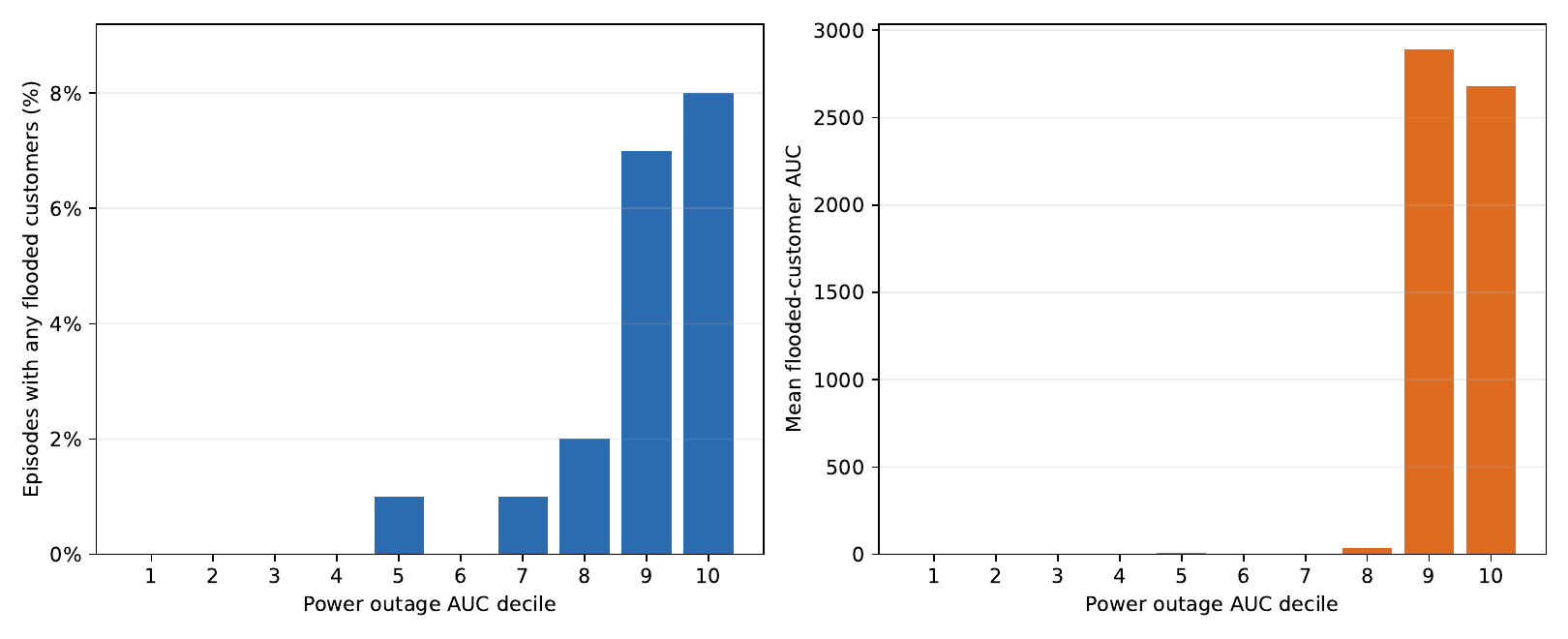}
\caption{Flood impacts in the selected coupled wind run grouped by power-outage-intensity decile.}
\label{fig:flood_coupling_severity}
\end{figure}

Figure~\ref{fig:flood_coupling_severity} reports the primary summary. Flood consequences concentrate in the severe power-outage tail. When episodes are grouped into deciles by power outage intensity, the probability of any flooded customers is 0\% through the first four deciles, rises to 1--2\% in the middle deciles, and reaches 7--8\% in the top two deciles. Mean flooded-customer intensity in the top power-outage-intensity decile is 2677, compared with an overall mean of 561 across all 1000 episodes. Among the 19 flood-positive episodes, the mean flooded-customer peak is 2727 customers, with a median of 974 and a p95 of 7389. These patterns show that both flood occurrence and flood intensity increase with outage intensity, which is consistent with the intended pump-failure-to-flood coupling logic rather than a trivial mirror of every power outage.

\subsection{Coarse external comparisons}
Table~\ref{tab:supplemental} summarizes coarse external comparisons against the two survey-based datasets. These numbers are not interpreted as pass/fail scores for the framework. The survey responses are temporally old, spatially shifted for privacy, and based on broad retrospective questions rather than event-aligned measurements. We therefore report these results only as coarse external context rather than as core evidence for the framework.

\begin{table}[H]
\centering
\caption{Survey-based external comparisons against the two supplemental datasets.}
\label{tab:supplemental}
\begin{tabular}{lll}
\toprule
Metric & Power survey & Flood survey \\
\midrule
Sample size & 42 & 41 \\
Weighted Pearson & 0.138 & 0.053 \\
Pearson & 0.122 & 0.006 \\
Spearman & 0.075 & 0.149 \\
Coverage & 613/613 & 2061/2061 \\
\bottomrule
\end{tabular}

\end{table}

\section{Discussion}
\label{sec:discussion}
The results suggest that the most decision-relevant outputs in this public-data setting are comparative rather than literal. The framework is most useful for asking how outage outcomes change under different fragility, topology, restoration, and repair-order assumptions while conditioning on historically plausible weather. Under that interpretation, the wind-event study shows that restoration duration and outage intensity are especially sensitive to topology and recovery-capacity assumptions, and the coupled flooding extension shows that pump-dependent flood consequences concentrate in the severe-outage tail rather than appearing uniformly across all outage episodes.

The same experiments also make clear where stronger claims are not yet justified. Historical fidelity is limited by the quality of the available observations and the missing utility-specific information behind them. Outage polygons are only a proxy for customer interruption, graph-defined customers are only a proxy for actual utility exposure, and the proxy network omits true feeder topology, underground/overhead labeling, asset condition, protection settings, and restoration practice during ongoing weather. These missing inputs limit how closely any public-data model can reproduce individual historical events, even when the weather forcing is event-specific.

\section{Conclusion}
\label{sec:conclusion}
This study shows that the framework can support resilience analysis conditioned on historical events under public-data constraints. Across five curated wind events, the simulated outage distributions stabilize at practical episode counts and respond coherently to fragility, topology, restoration, and repair assumptions. In the coupled wind run, flood consequences remain rare overall but concentrate in the severe-outage tail, indicating that pump-dependent flooding is a selective downstream consequence rather than a uniform byproduct of all outage episodes.

These results support using the framework primarily for comparative scenario analysis rather than literal reconstruction of individual historical outages. Under the current public-data assumptions, the most decision-relevant insights come from how outage duration, outage intensity, and coupled flooding change across modeling choices and historically conditioned weather events. Future work should prioritize utility-specific topology, underground/overhead labeling, outage records, and operational restoration data, while also improving winter-event modeling to support more realistic utility-facing studies.

\section*{Acknowledgments}
This material is based upon work supported by the Department of Energy, Solar Energy Technologies Office (SETO) Renewables Advancing Community Energy Resilience (RACER) program under Award Number DE-EE0010413. Any opinions, findings, conclusions, or recommendations expressed in this material are those of the authors and do not necessarily reflect the views of the Department of Energy.

\section*{Declaration of Generative AI and AI-assisted technologies in the writing process}
During the preparation of this work, the authors used OpenAI GPT-5.4 to assist with language editing, clarity improvement, and revision of manuscript text during the final drafting stage. The tool was used only to support writing and editing; all technical content, interpretations, and final wording were reviewed and approved by the authors, who take full responsibility for the content of the publication.

\bibliographystyle{elsarticle-num}
\bibliography{references}

\end{document}